\documentclass[11pt]{article}

\usepackage{color, cite}

\definecolor{rosso}{cmyk}{0,1,1,0.4}
\definecolor{rossos}{cmyk}{0,1,1,0.55}
\definecolor{rossoc}{cmyk}{0,0.5,1,0.2}
\definecolor{blu}{cmyk}{1,1,0,0.3}
\definecolor{blus}{cmyk}{1,1,0,0.6}
\definecolor{blucc}{cmyk}{1,0.4,0.2,0}
\definecolor{viola}{cmyk}{0,1,0,0.6}
\definecolor{viola2}{cmyk}{0,1,0.2,0.6}
\definecolor{verde}{cmyk}{0.92,0,0.59,0.25}
\definecolor{verdec}{cmyk}{0.92,0,0.59,0.15}
\definecolor{verdes}{cmyk}{0.92,0,0.59,0.4}
\font\tenrsfs=rsfs10 at 12pt
\font\sevenrsfs=rsfs7
\font\fiversfs=rsfs5
\newfam\rsfsfam

\textfont\rsfsfam=\tenrsfs
\scriptfont\rsfsfam=\sevenrsfs
\scriptscriptfont\rsfsfam=\fiversfs
\def\mathscr#1{{\fam\rsfsfam\relax#1}}

\def\circa#1{\,\raise.3ex\hbox{$#1$\kern-.75em\lower1ex\hbox{$\sim$}}\,}

\usepackage{amsmath,latexsym,amssymb,color,hyperref,graphicx}
\newcommand{\be}{\begin{equation}}
\newcommand{\ee}{\end{equation}}
\newcommand{\bea}{\begin{eqnarray}}
\newcommand{\eea}{\end{eqnarray}}
\newcommand{\no}{\noindent}
\newcommand{\nb}{\nonumber}

\renewcommand\b{\beta}

\renewcommand\L{\ensuremath{\Lambda_3}}

\newcommand{\ba}{\begin{eqnarray}}
\newcommand{\ea}{\end{eqnarray}}
\renewcommand\j{\ensuremath{\varphi}}
\newcommand{\B}{B}
\newcommand{\F}{{\cal F}}
\newcommand{\G}{{\cal G}}
\newcommand{\g}{\gamma}

\makeatletter
\def\ps@mine{%
    \def\@oddfoot{\hfil\thepage\hfil}\let\@evenfoot\@oddfoot
    \let\@oddhead\@evenhead%
    \let\@mkboth\@gobbletwo
    \let\sectionmark\@gobble
    \let\subsectionmark\@gobble
    }
\renewcommand\section{\@startsection {section}{1}{\z@}%
                                   {-3.5ex \@plus -1ex \@minus -.2ex}%
                                   {2ex \@plus.2ex}%
                                   {\normalfont\large\sffamily\bfseries}}
\renewcommand\subsection{\@startsection {subsection}{1}{\z@}%
                                   {-3.5ex \@plus -1ex \@minus -.2ex}%
                                   {2ex \@plus.2ex}%
                                   {\normalfont\sffamily\bfseries}}
\makeatother

\begin{document}

\def\FILL{\hfill\hfill\hfill}

\title{\sffamily\bfseries 
Self-accelerating universe in scalar-tensor theories \\ after GW170817   \\[3ex]
  \normalsize
   Marco Crisostomi$^a$ and Kazuya Koyama$^b$\\[2ex]
   \it\small
   $^a$Institut de physique th\'eorique, Universit\'e Paris Saclay CEA, CNRS, 91191 Gif-sur-Yvette, France \FILL\\
   Astrophysics Department, IRFU, CEA, Universit\'e Paris-Saclay, F-91191, Gif-sur-Yvette, France \FILL\\
   Laboratoire de Physique Th\'eorique, CNRS, Universit\'e Paris-Sud, Universit\'e Paris-Saclay, 91405 Orsay, France \FILL\\
   $^b$Institute of Cosmology and Gravitation, University of Portsmouth, Portsmouth, PO1 3FX, UK\FILL\\[2ex]
   {\tt marco.crisostomi@ipht.fr}, {\tt kazuya.koyama@port.ac.uk},\FILL\\[-6ex]
}

\date{\small \today}

\maketitle

\thispagestyle{empty}


\def\abstractname{\sc Abstract}
\begin{abstract}
\no The recent simultaneous detection of gravitational waves and a gamma ray burst from a neutron star merger significantly shrank the space of viable scalar-tensor theories by demanding that the speed of gravity is equal to that of light. The survived theories belong to the class of degenerate higher order scalar-tensor theories. We study whether these theories are suitable as dark energy candidates. We find scaling solutions in the matter dominated universe that lead to de Sitter solutions at late times without the cosmological constant, realising self-acceleration. 
We evaluate quasi-static perturbations around self-accelerating solutions and show that the stringent constraints coming from astrophysical objects and gravitational waves can be satisfied, leaving interesting possibilities to test these theories by cosmological observations. 
\end{abstract}

\bigskip

\section{Introduction}

In recent years, scalar-tensor theories of gravity have played a prominent role as modified gravity candidates, or dark energy models, to explain the observed accelerated expansion of our universe \cite{Clifton:2011jh, Joyce:2014kja, Koyama:2015vza}. In this context several interesting models have been developed including galileons \cite{Nicolis:2008in} and its covariantised version \cite{Deffayet0}. Horndeski theory \cite{Horndeski:1974wa} has been rediscovered \cite{Deffayet1, Deffayet2, Kobayashi:2011nu}, which provides the most general scalar-tensor theory in four dimensions leading to second order field equations. Within this theory, it is possible to obtain accelerated expansions of the universe without introducing the cosmological constant (self-acceleration) \cite{Chow:2009fm,Silva:2009km,Deffayet:2010qz, Kimura:2010di}.

The almost simultaneous detection of gravitational waves and gamma-ray bursts from the merging of a neutron stars binary system \cite{TheLIGOScientific:2017qsa, Coulter:2017wya, GBM:2017lvd,Murguia-Berthier:2017kkn} unequivocally fixed the speed of gravity $c_{GW}$ to be the same as the speed of light $c_{\text{light}}$ today to an accuracy of $10^{-15}$ \cite{Monitor:2017mdv}. This had a strong impact on the Horndeski theory \cite{Lombriser:2015sxa, Lombriser:2016yzn, Bettoni:2016mij,Creminelli:2017sry, Sakstein:2017xjx, Ezquiaga:2017ekz, Baker:2017hug, Arai:2017hxj}. Indeed the survived terms in the Horndeski action simply reduce to well known theories, K-essence, generalised Brans-Dicke and generalized cubic galileon, if we demand that the condition  $c_{GW}=c_{\text{light}}$ holds always.  

The fate of scalar-tensor theories however, does not need to coincide with that of Horndeski theory. It has been realised that it is indeed possible to go beyond this theory and have higher order field equations without introducing Ostrogradsky instabilities \cite{Motohashi:2014opa, Motohashi:2016ftl, Klein:2016aiq, Crisostomi:2017aim, Motohashi:2017eya}. The so called beyond Horndeski theories proposed in \cite{Gleyzes:2014dya, Gleyzes:2014qga} (see also \cite{Zumalacarregui:2013pma}), found a final collocation inside the class of degenerate higher order scalar-tensor theories introduced in \cite{Langlois:2015cwa} and further analysed in \cite{Langlois:2015skt, Crisostomi:2016tcp, Crisostomi:2016czh, Achour:2016rkg, deRham:2016wji, BenAchour:2016fzp}. These theories, fully classified in \cite{BenAchour:2016fzp}, represent so far the most general scalar tensor theories that propagate three degrees of freedom\footnote{Recently, new chiral scalar-tensor theories which break parity in the gravity sector have been introduced \cite{Crisostomi:2017ugk}, however they can make sense only as Lorentz breaking theories.}.

Inside this much larger class, the gravitational wave constraint $c_{GW}=c_{\text{light}}$ still leaves interesting non-trivial theories. It is therefore very important to understand whether these models can provide self-acceleration, i.e. an alternative to the $\Lambda$CDM model, to explain the accelerated expansion of the universe without cosmological constant. In this paper we undertake this analysis. 

The structure of the paper is as follows. In section 2, we introduce the degenerate higher order scalar tensor theories that satisfy the condition $c_{GW}=c_{\text{light}}$. In section 3, the equations to describe background cosmology as well as non-linear quasi-static perturbations are derived. In section 4, after showing that self-accelerating solutions indeed exist, we show an example of the scaling solution in the matter dominated universe that leads to the de Sitter solution at late time. We then study the predictions for linear and non-linear quasi-static perturbations in this background cosmology, demonstrating characteristic features of these models compared to $\Lambda$CDM. We also show how these models can pass the stringent constraints coming from astrophysical objects and gravitational waves. Section 5 is devoted to conclusions.

\section{Scalar tensor theory}
We consider the most general Lagrangian quadratic in second derivatives of the scalar field\footnote{The theories involving also cubic terms are completely excluded by the gravitational waves constraint $c_{GW}=c_{\text{light}}$.}
\cite{Langlois:2015cwa, Crisostomi:2016czh, Achour:2016rkg}
\be
{\cal L}_{quad} \,=\,\sum_{i=1}^5{\cal L}_i+{\cal L}_R \,, \label{ESTlag}
\ee
where
\begin{align}
{\cal L}_1 [A_1] &= A_1(\phi,\,X) \phi_{\mu \nu} \phi^{\mu \nu} \,,
\label{A1}
\\
{\cal L}_2 [A_2] &= A_2(\phi,\,X) (\Box \phi)^2 \,, 
\label{A2}
\\
{\cal L}_3 [A_3] &= A_3(\phi,\,X) (\Box \phi) \phi^{\mu} \phi_{\mu \nu} \phi^{\nu} \,, 
\label{A3}
\\
{\cal L}_4 [A_4] &= A_4(\phi,\,X)  \phi^{\mu} \phi_{\mu \rho} \phi^{\rho \nu} \phi_{\nu} \,, 
\label{A4}
\\
{\cal L}_5 [A_5] &= A_5(\phi,\,X)  (\phi^{\mu} \phi_{\mu \nu} \phi^{\nu})^2\,, 
\label{A5}
\end{align}
while
\begin{equation}
{\cal L}_R[G] = G(\phi,\,X) R \,,
\label{R}
\end{equation}
is a non-minimal coupling with gravity. We defined $\phi_{\mu} = \nabla_{\mu} \phi, \phi_{\mu \nu} = \nabla_{\mu} \nabla_{\nu} \phi$ and $X = \phi^{\mu} \phi_{\mu}$.  The functions $G$, $A_i$ are arbitrary functions of $\phi$ and  $X$ but for simplicity we will only consider them to be functions of $X$. For the Lagrangian (\ref{ESTlag}), the speed of gravitational waves computed from linear tensor perturbations around the cosmological background was firstly given in \cite{deRham:2016wji}:
\begin{equation}
c_{GW}^2 = \frac{G}{G -  X A_1}, \label{GWcond}
\end{equation}
in the units where $c_{\text{light}}=1$. 

The Horndeski theory is given by the following choice of functions
\begin{equation}
A_2=-A_1= -2 G_{4 X}\,, \qquad A_3=A_4=A_5=0,
\label{HOR}
\end{equation}
where $G_{4X} = d G_4/dX$. Therefore, Eq.~(\ref{GWcond}) implies that $G_{4X}$ needs to be tuned to be small at least at the vicinity ($<40$ Mpc) of the Solar System today ($z<0.01$) \cite{Lombriser:2015sxa, Lombriser:2016yzn, Bettoni:2016mij, Creminelli:2017sry, Sakstein:2017xjx, Ezquiaga:2017ekz, Baker:2017hug, Arai:2017hxj}. In this paper, we do not consider this tuning. 

For the general Lagrangian (\ref{ESTlag}), Eq.~(\ref{GWcond}) simply implies that the condition  
\be
A_1=0 \,,
\ee
needs to be satisfied to ensure $c_{GW}=1$, and this is what we will assume in this paper.   
To satisfy the degeneracy conditions that remove the Ostrogradsky ghost, the other functions should satisfy the following relations
\bea
\label{degcond}
A_2&=&0\,, \qquad A_5 = 
\frac{A_3}{2 G}(4 G_X + A_3 X)\,, \\
A_4&=& -\frac{1}{8 G}\left[
8 A_3 G - 48 G_{X}^2 - 8 A_3 G_{X} X + A_3^2 X^2
\right] \,, \nonumber
\eea
whereas $G$ and $A_3$ are left free.
This theory, with two free functions, is a subset of the class called N-I in \cite{Crisostomi:2016czh} and Ia in \cite{Achour:2016rkg} with $A_1=0$. 

We assume that matter is minimally coupled to the metric~$g_{\mu \nu}$. Finally, we can also add lower order Horndeski Lagrangians to (\ref{ESTlag}) without spoiling the condition $c_{GW}=1$. The total Lagrangian is given by 
\begin{align}
{\cal L}_{\text{tot}}= G_2(X) + G_3(X)\Box \phi + {\cal L}_{\text{quad}}
+{\cal L}_{\text{m}} [g_{\mu \nu}] \,,
\end{align}
where again we will assume that $G_2$ and $G_3$ are functions of $X$ only.

Beyond Horndeski theory \cite{Gleyzes:2014dya, Gleyzes:2014qga} is contained in the above theory and corresponds to the following choice of the two free functions
\be
A_3 = - 4 G_{X}/X \,.
\label{HbH}
\ee
In order to underline the differences with beyond Horndeski theory, we introduce a new function $\B_1$ defined as
\be
\B_1 \equiv \frac{X}{4 G} (4 G_X + X A_3 ) \,.
\label{beta1}
\ee
This $B_1(X)$ is related to the parameter $\beta_1$ introduced in the framework of the effective field theory of dark energy \cite{Langlois:2017mxy}. The beyond Horndeski limit therefore can be easily obtained by sending $\B_1$ to zero. We will replace $A_3(X)$ with $\B_1(X)$ in the rest of the paper.

\section{Cosmological background, linear and non-linear quasi-static perturbations}

We consider a cosmological background with a time dependent scalar field $\phi=\j(t)$ and study the perturbations around it, namely
\begin{equation}
ds^2 = -(1+ 2 \Phi(t, x^i)) dt^2 + a(t)^2 (1- 2 \Psi(t, x^i)) 
\delta_{ij} dx^i dx^j,
\end{equation}
with $\phi = \varphi(t) + \pi(t,x^i)$. The Hubble function is defined as $H(t)= \dot{a}/a$. Since we are interested in the application of this theory to the later time universe, we only consider matter with no pressure and define the matter density as $\rho(t) + \delta\rho(t,x^i)$.

\medskip

At the background level, the $tt$ component of the Einstein equation contains $\dot{H}$ and $\dddot{\phi}$, and the scalar field equation contains $\ddddot{\phi}$,  $\dddot{\phi}$, $\ddot{H}$ and $\dot{H}$. To obtain second order equations written only with $H$, $\dot{\phi}$ and $\ddot{\phi}$, we use the $ii$ equation to determine $\dot H$ and substitute it and its time derivative in the $tt$ and scalar field equation. We then get the modified Freedman equation and the scalar field equation
\bea
&& {\cal A} (\dot{\phi}, \ddot{\phi}) H^2 + {\cal B} (\dot{\phi}, \ddot{\phi})H + {\cal C}(\dot{\phi}, \ddot{\phi}) = \rho \,, \label{back1} \\
&& {\cal D}(\dot{\phi}, \ddot{\phi}) H^2 + {\cal E} (\dot{\phi}, \ddot{\phi})H + {\cal F}(\dot{\phi}, \ddot{\phi}) = 0 \,, \label{back2}
\eea
where the above coefficients are given in Appendix \ref{background}. We can further solve $H$ in terms of $\dot{\phi}$ and $\ddot{\phi}$ using Eq.~(\ref{back1}) and substitute it to Eq.~(\ref{back2}) to obtain the second order equation for the scalar field only. 

\medskip

Beyond the linear terms in perturbations, to identify the relevant non-linear terms describing the Vainshtein mechanism, we expand the equations of motion in terms of the fluctuations using the following assumptions \cite{Kimura:2011dc, Koyama:2013paa, Kobayashi:2014ida}: the fields $\pi$, $\Phi$ and $\Psi$ are small, hence we neglect higher order interactions containing the metric perturbations $\Phi$ and $\Psi$, as well as terms containing higher order powers of the scalar field fluctuation $\pi$ and its first derivatives. On the other hand, we keep all terms with second or higher order spatial derivatives of perturbations. 

We will also work with quasi-static approximations and ignore the time derivatives of the perturbations compared with the spatial derivatives. Note that we need to keep time derivatives for the terms containing second or higher order spatial derivatives in order to be consistent with the expansion scheme.

With these assumptions, we obtain the following equations describing the dynamics of these perturbations \cite{Crisostomi:2017lbg}:
\begin{align}
{\cal G}_T \nabla^2 \Psi + {\cal G}_{T \Phi} \nabla^2 \Phi + a_2 \nabla^2 \pi
+a_{2}^t \nabla^2 \dot{\pi}
+ b_2^{a} (\nabla^2 \pi)^2 + b_2^{b} (\nabla_{i} \nabla_j \pi)^2
+b_2^{c} (\nabla^i \pi)(\nabla_i \nabla^2 \pi) = a^2 \rho \, \delta, \label{eqtt}
\end{align}
\begin{align}
{\cal F}_T \nabla^2 \Psi - {\cal G}_T \nabla^2 \Phi + a_1^t \nabla^2 \dot{\pi}
+ b_1 (\nabla_i \nabla_j \pi)^2 
+b_1 (\nabla^i \pi)(\nabla_i \nabla^2 \pi) =0, \label{eqii}
\end{align}
\begin{align}
&a_0 \nabla^2 \pi + a_0^t \nabla^2 \dot{\pi} + a_0^{tt} \nabla^2 \ddot{\pi} 
+a_1 \nabla^2 \Psi + 2 a_1^t \nabla^2 \dot{\Psi} 
+a_3 \nabla^2 \Phi - 2 a_2^t \nabla^2 \dot{\Phi} \nonumber\\
&+ b_0^a (\nabla^2 \pi)^2 
+b_0^{b} (\nabla_i \nabla_j \pi)^2
+ b_0^{c} (\nabla^i \pi)(\nabla_i \nabla^2 \pi) \nb \\
&+b_0^t (\nabla^2 \dot{\pi} )(\nabla^2 \pi) 
+ 2 b_0^t (\nabla_i \nabla_j \dot{\pi}) (\nabla^i \nabla^j \pi) 
+b_0^t (\nabla^i \dot{\pi}) (\nabla_i \nabla^2 \pi)
+2 b_0^t (\nabla^i \pi)(\nabla_i \nabla^2 \dot{\pi}) \nb \\
& + 2 b_1 (\nabla^2 \Psi)(\nabla^2 \pi) + 2 b_1
(\nabla^i \nabla^2 \Psi) (\nabla_i \pi)  
-4 b_2^a 
(\nabla_i \nabla_j \Phi) (\nabla^i \nabla^j \pi)
- 2 b_2^c
(\nabla^i \nabla^2 \Phi) (\nabla_i \pi) + b_3 (\nabla^2 \Phi)(\nabla^2 \pi) \nonumber\\
&+ c_0^a (\nabla^2 \pi)^3 + 2 c_0^a (\nabla_i \nabla_j \pi)^3 + c_0^b (\nabla_i \nabla_j \pi)^2 (\nabla^2 \pi) 
+ c_0^c (\nabla^i \pi)(\nabla_i \nabla_j \pi)(\nabla^j \nabla^2 \pi)   \nonumber\\
&+ c_0^c (\nabla_i \pi) 
(\nabla^i \nabla^2 \pi) (\nabla^2 \pi)
+ 2 c_0^c (\nabla^i \pi)(\nabla_i \nabla_k \nabla_j \pi)
(\nabla^k \nabla^j \pi)
+ c_0^c (\nabla^i \pi)(\nabla^j \pi)(\nabla^2
\nabla_i \nabla_j \pi)
 =0 \,, \label{eqss} 
\end{align}
where $\nabla_i$ is the spatial derivative with respect to $\delta_{ij}$. We give the explicit expression of these coefficients in Appendix \ref{perturbations}.
We defined the energy density contrast as $\delta \equiv \delta\rho/\rho$ and, under this approximation,
the continuity equation for matter and the Euler equation read
\begin{equation}
\dot{\delta}= - \frac{\nabla^2 v}{a^2} \,, \qquad \dot{v}= - \Phi \,,
\label{continuity}
\end{equation}
where $v$ is the velocity potential.

\medskip

At the linear order, Eqs.~(\ref{eqtt}) and (\ref{eqii}) can be solved to determine $\nabla^2 \Phi$ and $\nabla^2 \Phi$ in terms of the scalar field $\pi$. Substituting these expressions and their time derivatives in the linear scalar field equation (\ref{eqss}), we get a second order equation for $\pi$ only, of the form
\be
\gamma_1 \nabla^2 \pi + \gamma_2 a^2 \rho \, \delta + \gamma_3 \rho \nabla^2 v = 0 \,,
\ee
where $\gamma_{1,2,3}$ are given in Appendix \ref{linear}, and
we used the continuity equation (\ref{continuity}) to replace $\dot{\delta}$ by $v$.  

Solving the above equation for $\nabla^2 \pi$ and substituting it and its time derivatives in the linear (\ref{eqtt}) and (\ref{eqii}) equations, we find the expression for the metric potentials
\bea
&& \nabla^2 \Phi = \mu_\Phi a^2 \rho \, \delta + \nu_\Phi \rho \nabla^2 v \,, \\
&& \nabla^2 \Psi = \mu_\Psi a^2 \rho \, \delta + \nu_\Psi \rho \nabla^2 v \,,
\eea
where the coefficients $\mu$ and $\nu$ are given in Appendix \ref{linear}. Here we used the Euler equation (\ref{continuity}) for matter perturbations.

Finally, using the conservation equations for matter perturbations (\ref{continuity}), we get the evolution equation for the linear density perturbation
\be
\ddot\delta + (2H+ \nu_\Phi \rho) \dot\delta - \mu_\Phi \rho\,\delta = 0. 
\ee
As already discussed in the context of the effective field theory of dark energy
\cite{DAmico:2016ntq}, there appear two modifications. One is the correction to the effective Newton constant $\mu_{\Phi}$, and the other is an extra damping term $\nu_{\Phi}$. This damping term is absent in the Horndeski theory. If $\nu_{\Phi} >0$, the additional damping could suppress the growth of the structure.  

\medskip

On small scales, we need the Vainshtein mechanism to hide the modification and satisfy the stringent Solar System constraints. To identify on which scale the non-linear terms become important, we introduce a mass dimension $\L$ and assume the following scaling for the functions $G, A_3, A_4$ and $A_5$
\be
G \sim M_p^{2}, \;  X A_3 \sim X A_4 \sim X^2 A_5
\sim M_{p}  \L^{-3},
\label{scaling}
\ee
where $M_p$ is the Planck mass and we assume $X \sim M_p \L^3$. If the background scalar field is responsible for dark energy, then we expect $\L^3 \sim H_0^2 M_P$ where $H_0$ is the present-day Hubble parameter.  In \cite{Crisostomi:2017lbg} (see also \cite{Langlois:2017dyl}) we computed the spherically symmetric solutions and found that the non-linear terms are important below the Vainshtein radius defined by $r{_V} =  (M/M_p \L^3)^{1/3} $. For $r \ll r_V$
the following solutions for metric perturbations were found
\begin{align}
\Phi' & = \frac{G_N M}{r^2} + \frac{\Upsilon_1 G_N}{4} M'', \nonumber\\
\Psi' & = \frac{G_N M}{r^2} - \frac{5 \Upsilon_2 G_N}{4 r} M' 
+ \Upsilon_3 G_N M'',
\label{spherical}
\end{align}
where 
\begin{align}
\Upsilon_1 & = -\frac{\left(2 \dot\j^2
	G_X+G \B_1\right)^2}{G \left(\dot\j^2
	G_X+G \B_1\right)} \,, \nonumber\\
\Upsilon_2 &=   \frac{8 G_{X} X}{5 G} \,,  \nonumber\\
\Upsilon_3 &= -\frac{\B_1 \left(2 \dot\j^2
	G_X+G \B_1\right)}{4 \left(\dot\j^2 G_X+G \B_1\right)} \,,
\label{Ypsilon}
\end{align}	
\begin{align}
G_N= \frac{1}{8\pi}\left[2 G \left(1 - 3 \B_1\right) - 4 \dot\j^2 G_X\right]^{-1} \,,
\label{GN}
\end{align}
and $X, G, G_{X}$ and $\B_1$ are all evaluated at the background.
\be
M(t,r) = \int_0^r 4 \pi r'^2 a(t)^2 \delta \rho(r',t) dr' \,,
\ee
is the enclosed mass within the radius $r$. 
Outside a matter source ($M'=M''=0$), the solutions (\ref{spherical}) reduce to those in GR with a time dependent Newton constant $G_N$, thus the Vainshtein mechanism is working to suppress the contributions from the scalar field perturbations. On the other hand, inside the matter source, the Vainshtein mechanism is broken and gravity is modified from GR, which leaves interesting signatures inside astrophysical objects. Note that $G_N$ is the Newton constant measured locally inside the Vainshtein radius and it is in general time dependent in this theory. 

\section{Self-accelerating solutions}
In the Horndeski theory, it is possible to obtain accelerated expansions of the universe without introducing the cosmological constant \cite{Chow:2009fm,Silva:2009km,Deffayet:2010qz, Kimura:2010di}. A typical example is the covariant galileon model \cite{Deffayet0}, in which the Horndeski functions are given by 
\be
G_2 = c_2 X \,, \quad G_3 = \frac{c_3}{\L^3}X \,, \quad G_4 = \frac{M_P^2}{2}+\frac{c_4}{\L^6} X^2  \,, \label{functs}
\ee
with a possible addition of the quintic Horndeski term with a parameter $c_5$. The interesting feature of this model is that it admits the scaling solution, $\dot{\phi} = \xi/H$, where $\xi$ is constant throughout the history of the universe \cite{DeFelice:2010pv, DeFelice:2010nf}. For appropriate choices of the parameters, this is an attractor solution, i.e. independently of initial conditions, the solution approaches this scaling solution. This allows us to constrain the model parameters $c_i$ independently of initial conditions.
Significant efforts have been made in the study of linear perturbations 
\cite{DeFelice:2010as, Barreira:2012kk, Appleby:2012ba, Appleby:2011aa}, non-linear structure formation \cite{Barreira:2013eea, Li:2013tda} and observational constraints \cite{Barreira:2013jma, Barreira:2014jha}. Unfortunately this model is excluded by the condition $c_{GW}=1$ today if $c_4 \neq 0$ or $c_5 \neq 0$, while the cubic galileon model with $c_4=c_5=0$ is excluded by cosmological observations \cite{Renk:2017rzu, Peirone:2017vcq}. Thus it is important to ask whether there are similar scaling solutions in the most general theory with $c_{GW}=1$.   

Following the covariant galileon, we choose our free functions as follows
\be
G_2 = c_2 X \,, \quad G_3 = \frac{c_3}{\L^3}X \,, \quad G = \frac{M_P^2}{2}+\frac{c_4}{\L^6} X^2 \,, \quad A_3 = - \frac{8 c_4}{\L^6} - \frac{\b}{\L^6} \,, \label{functs}
\ee
which implies $\B_1 = - \b X^2/\left(2 \L^6 M_p^2 + 4 c_4 X^2 \right)$.  Note that we have the same number of parameters as the covariant galileon model\footnote{It would be interesting to understand whether this choice of parameters still enjoys the weakly broken galileon symmetry as it happens for the very same choice of $G_2, G_3$ and $G$ in the Horndeski theory \cite{Pirtskhalava:2015nla}.}.

To make the equations dimensionless, we introduce the following dimensionless quantities: $t \to M_P^{1/2} \L^{-3/2} t, \; H \to M_P^{-1/2} \L^{3/2} H$ and $ \phi \to M_P \phi$, and work with them in the following.

We now seek the scaling solution of the form $\dot{\phi} =\xi/H$. Unlike the covariant galileon case, the Friedmann equation and the scalar field equation, Eqs.~(\ref{back1}) and (\ref{back2}),  contain $\ddot{\phi}$, which prevents us from finding scaling solutions in the entire history of the universe. However, it is still possible to find scaling solutions both in the Matter Dominated (MD) era and late time de Sitter (dS) phase.

During the MD era, we expect that the standard Friedman equation holds if the scaling solution exists as $\dot{\phi} = \xi/H \to 0$, thus $G \to M_P^2/2$ and other functions are negligible. Under this assumption we have $\dot{\phi} =3 \xi t/2$ and $\ddot{\phi} =3 \xi/2$. Therefore, we can ignore $\dot{\phi}$ while $\ddot{\phi}$ approaches a constant. Using this approximation in Eqs.~(\ref{back1}) and (\ref{back2}), we obtain the following solution in the MD era
\be
H \simeq H_{M} = \frac{2}{3 t} \,, \qquad \xi_M \simeq \frac{6 c_3-2 \sqrt{-9 \beta  c_2-48 c_2
   c_4+9 c_3^2}}{9 \beta +48 c_4} \,.
\ee

On the other hand, at late times, if the universe approaches de Sitter solution, $H$ becomes constant and if the scaling solution holds, $\dot{\phi}=\xi/H =$ const. and $\ddot{\phi}=0$. Using this approximation in Eqs.~(\ref{back1}) and (\ref{back2}), we obtain the following solution
\bea
&& H_{dS}^4 \simeq \frac{\left(\sqrt{9 c_3^2-6 c_2 (3 \beta +8
   c_4)}-3 c_3\right)^2 \left(c_2-\frac{2
   c_4 \left(\sqrt{9 c_3^2-6 c_2 (3 \beta +8
   c_4)}-3 c_3\right)^2}{3 (3 \beta +8
   c_4)^2}\right)}{3 (9 \beta +24 c_4)^2} \,, \\
&& \xi_{dS} \simeq \frac{3 c_3-\sqrt{9 c_3^2-6 c_2 (3 \beta +8
   c_4)}}{9 \beta +24 c_4}.
\eea
This realises the self-accelerating solution as in covariant galileon models. We notice that for $\beta=0$, $\xi_M = \xi_{dS}$ thus it is possible to find the scaling solution in the entire history of the universe. In fact, it was shown that the covariant galileon model can be extended to beyond Horndeski theory and the background solution remains the same although the perturbations behave differently \cite{Kase:2014yya}.

\begin{figure}[h]
	\begin{center}
		\includegraphics[width=8cm]{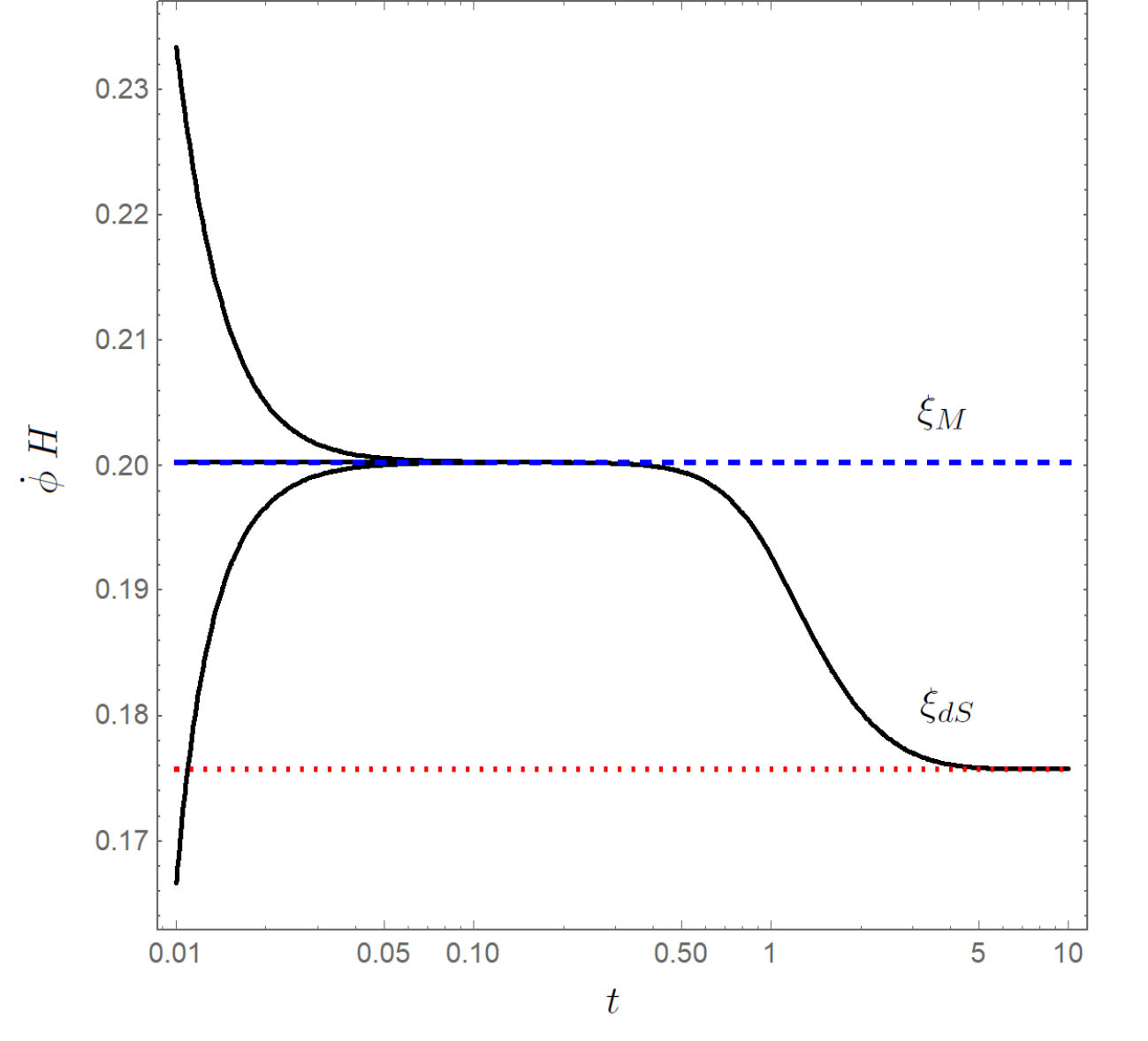}
	   \includegraphics[width=8cm]{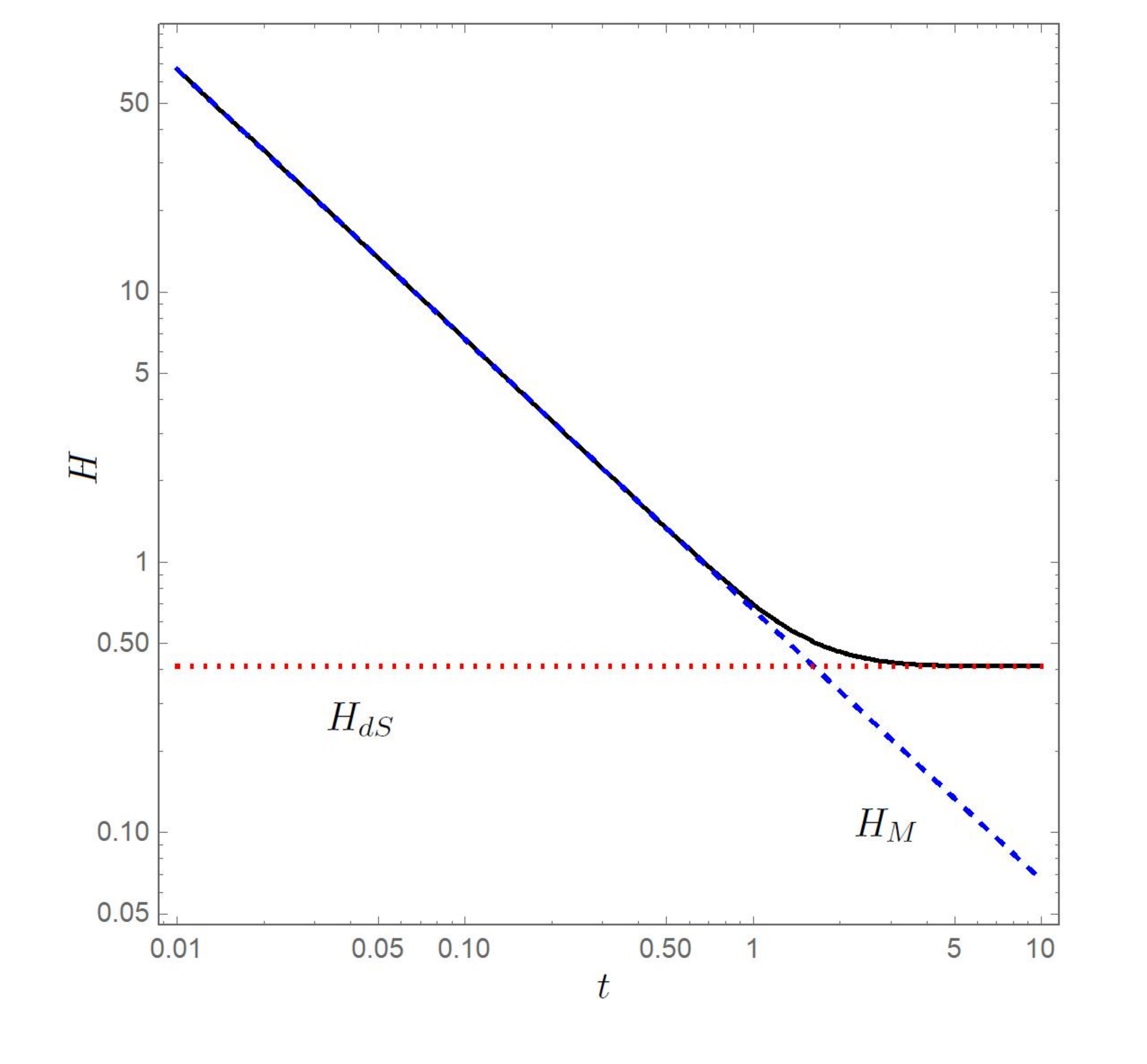}
		\caption{The evolution of $\dot{\phi}(t) H(t) $ and $H(t)$. The scalar field solution is attracted to the matter scaling solution first, and then to the de Sitter scaling solution. We show the evolutions with three different initial conditions. We choose the following parameters: $c_2=3, c_3=5, c_4=1, \beta=-5.3$ for this illustration.}
		\label{fig1}
	\end{center}
\end{figure}

Fig.~\ref{fig1} confirms these analytic solutions by numerically solving Eqs.~(\ref{back1}) and (\ref{back2}). The left panel shows $\dot{\phi} H$ with three different initial conditions. For this choice of parameters, the solutions are attracted to the MD scaling solution first. Then the solutions approach the dS scaling solution realising self-acceleration. In the right panel of Fig.~\ref{fig1} we plot $H(t)$ for the same choice of parameters together with $H_M(t)$ and $H_{dS}$. As it can be clearly seen, $H(t)$ follows $H_M(t)$ at early times and then approaches $H_{dS}$ when $t \sim 1$ (this corresponds to the dimension-full time $t \sim M_P^{1/2} \L^{-3/2}$).

\begin{figure}[h]
	\begin{center}
		\includegraphics[width=8cm]{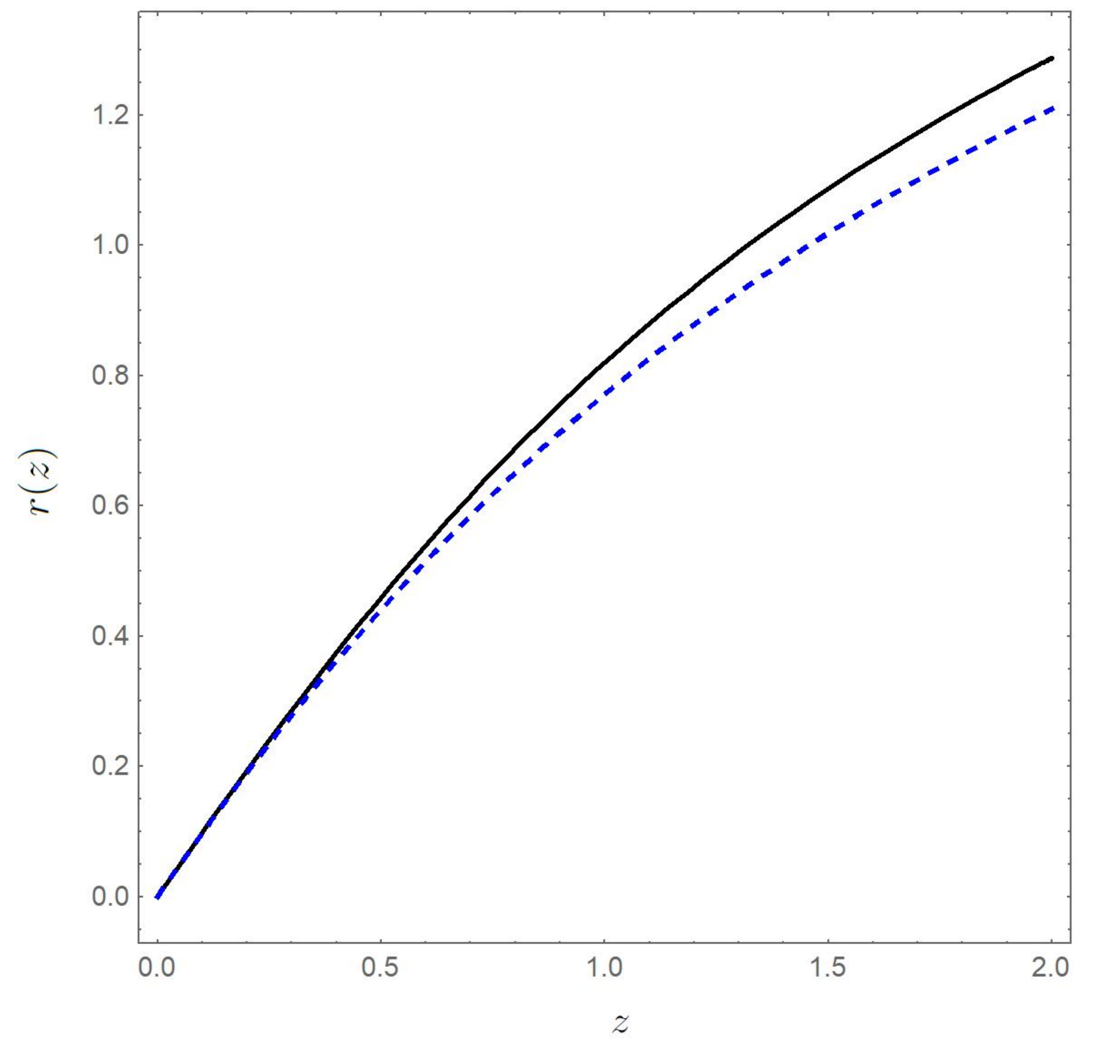}
		\includegraphics[width=8.3cm]{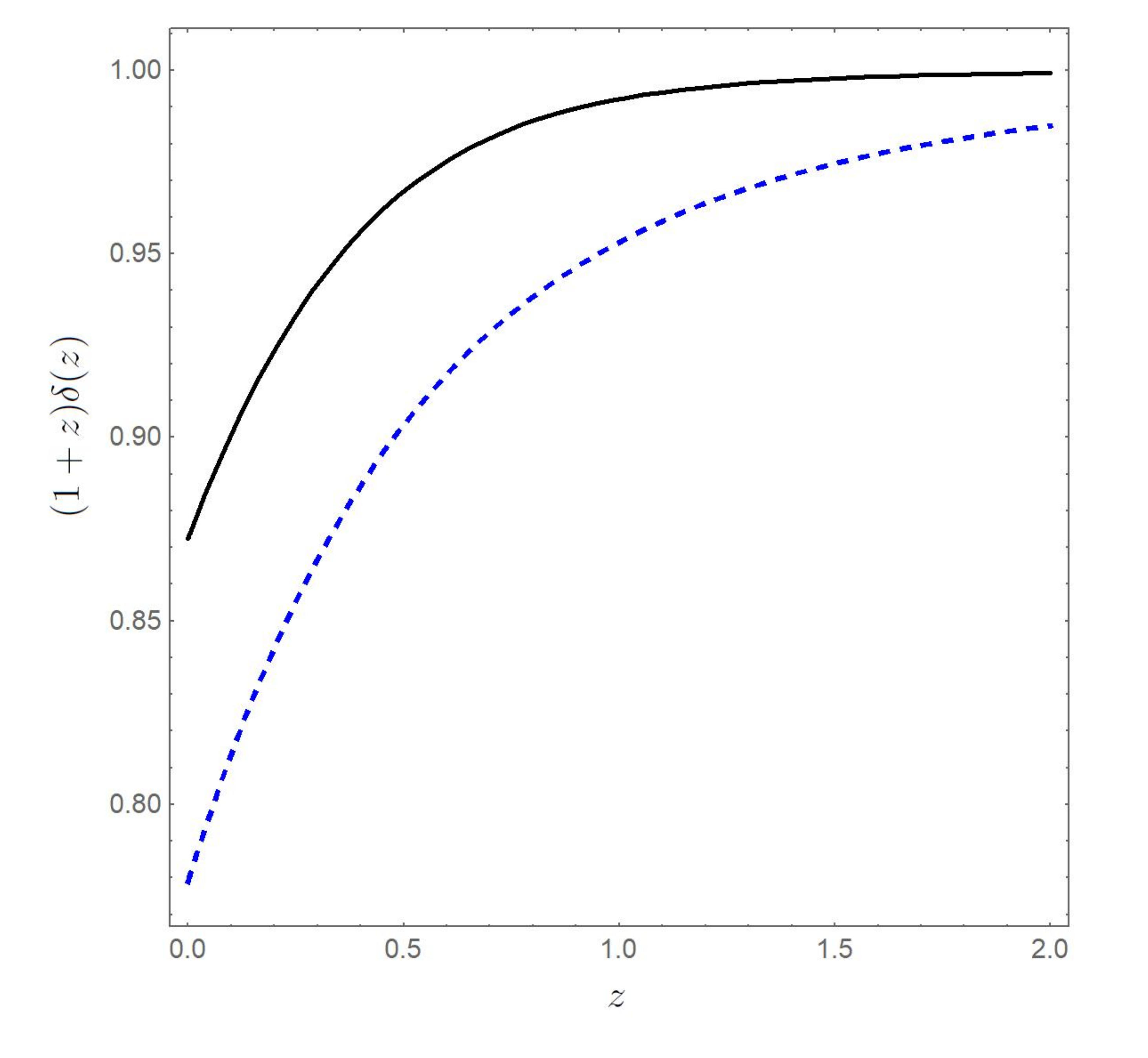}
		\caption{The comoving distance $r(z)$ and the density perturbation divided by the scale factor $(1+z) \delta(z)$. For a comparison, the $\Lambda$CDM prediction $\Omega_m=0,3$ is also shown (blue, dashed). The parameters are the same as Fig.~\ref{fig1}.}
		\label{fig2}
	\end{center}
\end{figure}

Given the background solution, we can evaluate the coefficients in the equations for quasi-static perturbations Eqs.~(\ref{eqtt}), (\ref{eqii}) and (\ref{eqss}), and find the solutions on this cosmological background. As an illustration, in Fig.~\ref{fig2}, we show the comoving distance $r(z) = \int_0^z dz'/H(z')$ in the left panel, and the evolution of density perturbations divided by the scale factor, $(1+z) \delta(z)$, in the right panel. As a comparison we show the $\Lambda$CDM result with $\Omega_m=0.3$. Note that we choose the time today $t_0$ to roughly match $H(z)$ with $H_{\Lambda CDM}(z)$ at high redshifts. In this example, the density perturbation is enhanced compared with $\Lambda$CDM. This is due to the enhanced effective Newton constant $\mu_{\Phi}>1$. In this example, the damping term $\nu_{\Phi}$ remains small and does not play a significant role. 

\begin{figure}[h]
	\begin{center}
		\includegraphics[width=8cm]{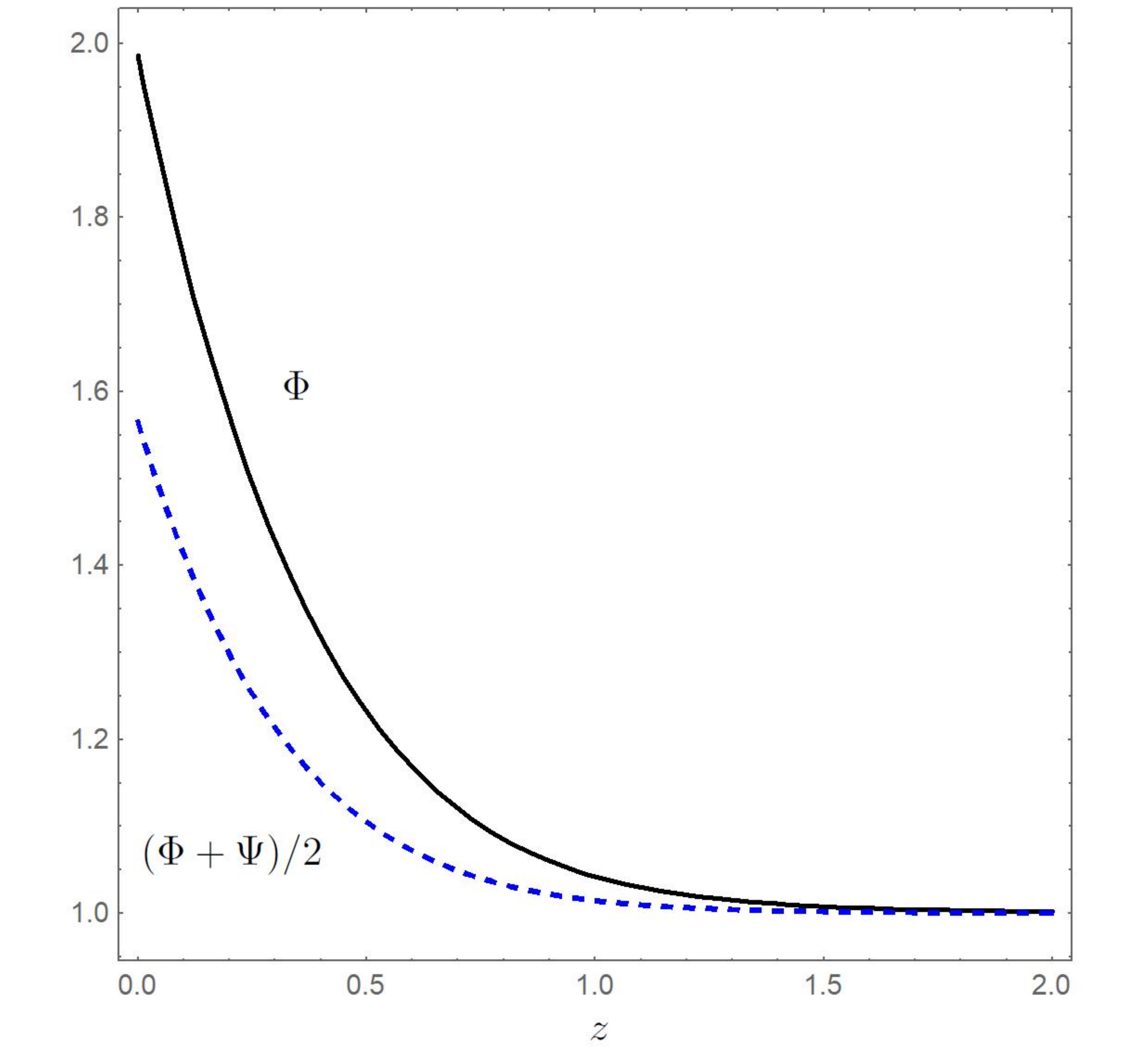}
		\includegraphics[width=8cm]{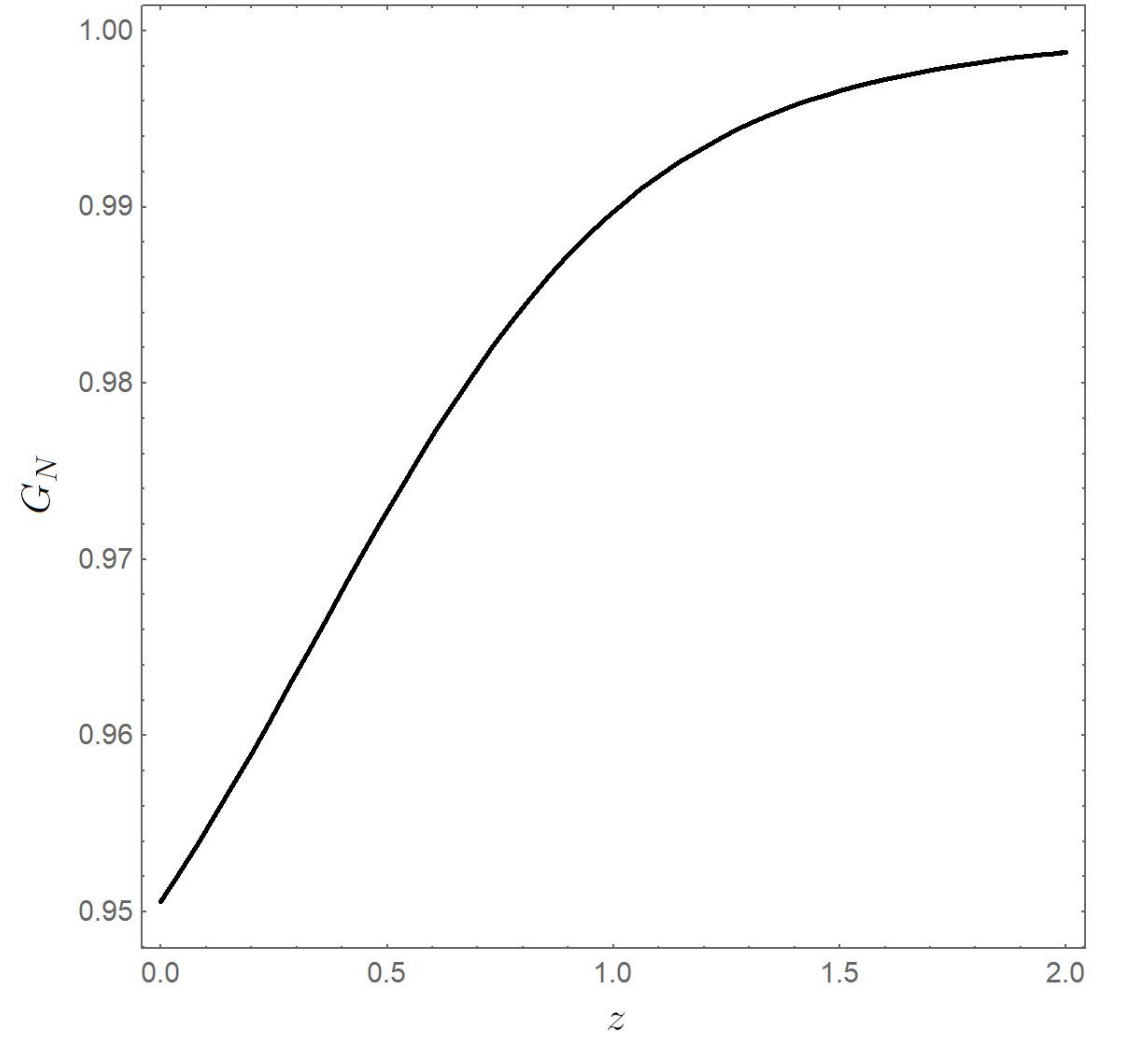}
		\caption{Left panel: the evolution of the Newtonian potential $\Phi(z)$ (black, solid) and the lensing potential $(\Phi(z)+\Psi(z))/2$ (blue dashed). The right panel shows the time dependence of the local Newton constant  $8 \pi M_P^2 G_N(z)$. The parameters are the same as Fig.~\ref{fig1}.}
		\label{fig3}
	\end{center}
\end{figure}

The left panel of Fig.~\ref{fig3} shows the evolution of the Newtonian potential $\Phi(z)$  and the lensing potential $(\Phi(z)+ \Psi(z))/2$. These potentials are normalised to be one at early times. In this example, both potentials grow at late times. The growth of the lensing potential is a common feature of the galileon model and this is in fact the origin of strong observational constraints, as the increase of the lensing potential leads to the opposite sign for the Integrated Sach-Wolfe (ISW) effect compared with $\Lambda$CDM. This signals that the self-acceleration solutions in this theory could be also strongly constrained by the ISW-galaxy cross correlation. However, in order to determine whether this rules out the model, we need to scan the whole parameter space carefully. In fact, unlike the covariant galileon theories, the evolution equation for the density perturbation has the extra damping term $\nu_{\Phi}$, and depending on the choice of parameters, this could help suppress the growth of the structure and the lensing potential as it happens in the context of beyond Horndeski theories \cite{DAmico:2016ntq}. In the right panel of Fig.~\ref{fig3}, we show the time evolution of the Newton constant $8 \pi M_P^2 G_N(z)$ measured locally (\ref{GN}). Unlike the effective Newton constant for linear perturbations, this is suppressed today. The time variation of $G_N$ is strongly constrained by the Lunar Laser Ranging experiment \cite{Williams:2004qba}. In order to impose this constraint, a precise determination of $t_0$ is required by fitting cosmological parameters (see \cite{Barreira:2014jha} in the case of covariant galileon models).   

\begin{figure}[h]
	\begin{center}
		\includegraphics[width=8cm]{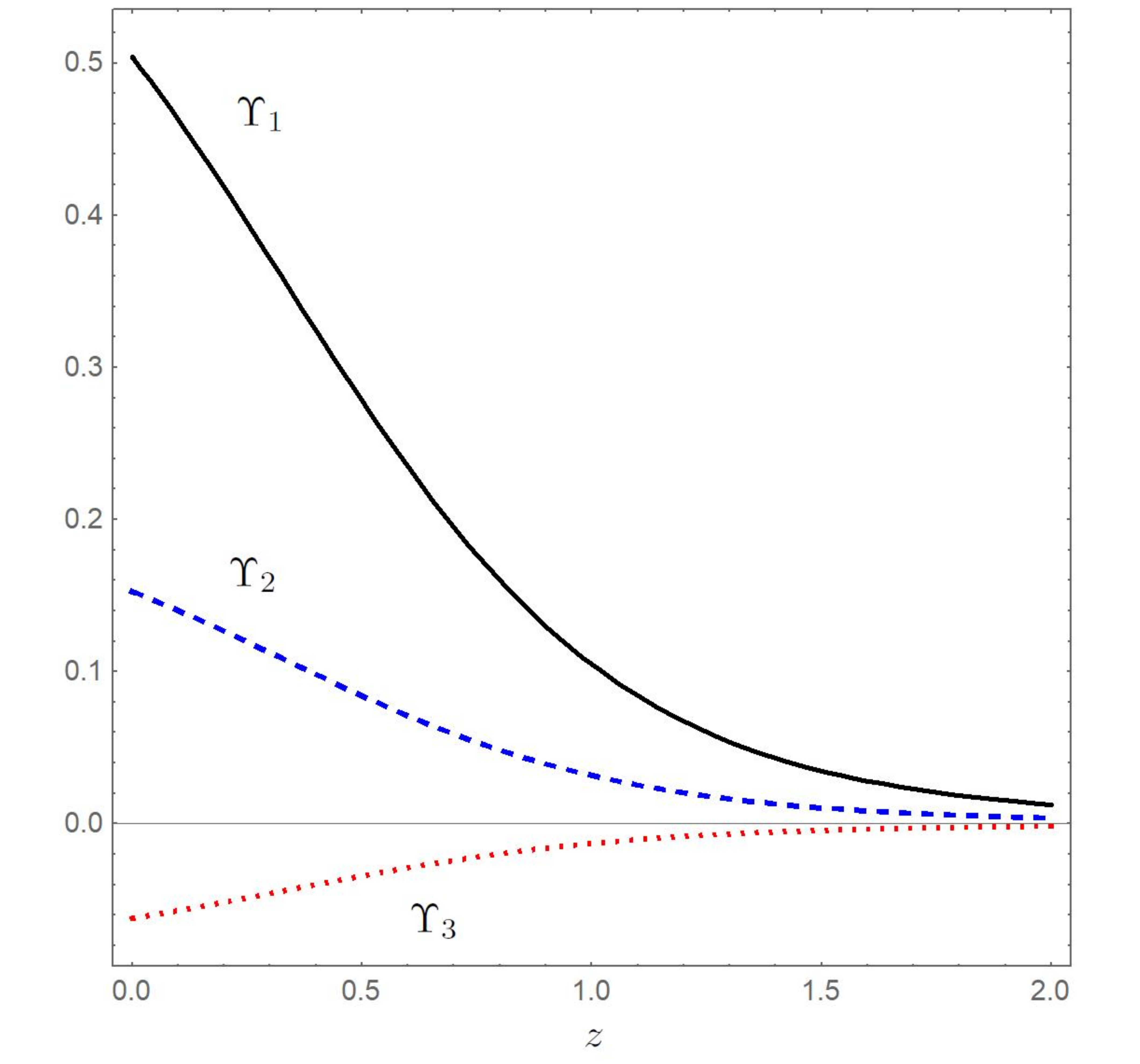}
		\includegraphics[width=8.8cm]{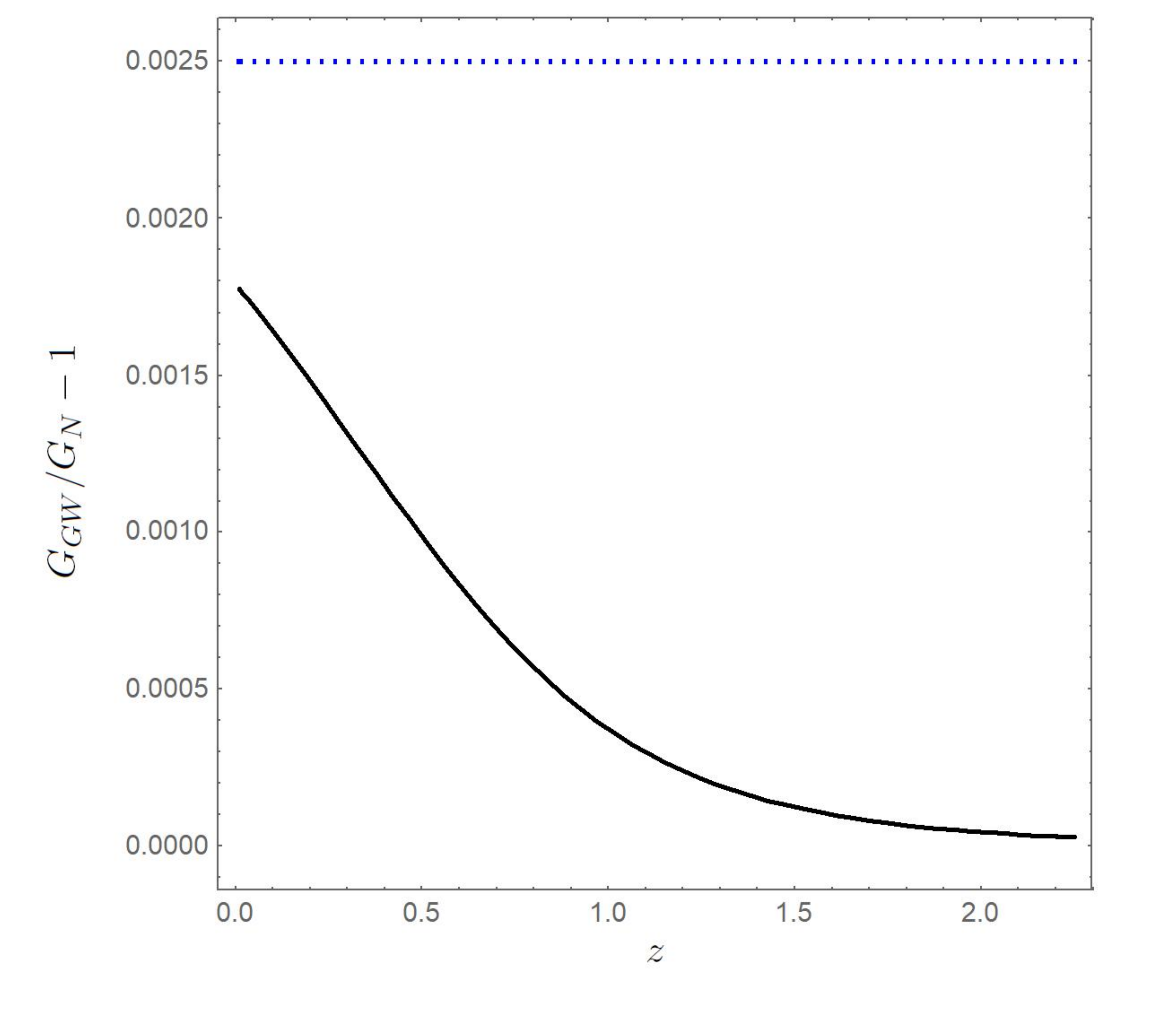}
		\caption{The left panel shows the evolution of $\Upsilon_i$ while the right panel shows $G_{GW}/G_{N}-1$. The blue dotted line on the right panel indicates the upper bound from the Hulse-Taylor binary. The parameters are the same as Fig.~\ref{fig1}.}
		\label{fig4}
	\end{center}
\end{figure}

Finally, in Fig.~\ref{fig4}, we show the coefficients that appear in the Vainshtein solutions for metric perturbations (\ref{Ypsilon}). In the left panel, we show the time evolution of $\Upsilon_i (z)$. $\Upsilon_1$ is constrained as $-2/3 < \Upsilon_1 < 1.6$ from the stellar structure \cite{Saito:2015fza, Sakstein:2015zoa, Sakstein:2015aac} and this model satisfies this condition. There is another tight constrain that comes from the orbital decay of Hulse-Taylor pulsar \cite{Jimenez:2015bwa}. The effective Newton constant for gravitational waves is given by $G_{GW}= 1/16 \pi G$, which is different from the Newton constant $G_N$ measured locally. The ratio between the two is well constrained \cite{Dima:2017pwp}
\be
-7.5 \times 10^{-3} < \frac{G_{GW}}{G_{N}} -1 < 2.5 \times  10^{-3}.
\ee
In this theory we have
\be
\frac{G_{GW}}{G_{N}}-1 =\frac{2 X G_{X}}{G} - 3 B_1(X).
\ee
In the right panel of Fig.~\ref{fig4} we plot $G_{GW}/G_{N}-1$ for this model showing that this constraint is also satisfied.

\section{Conclusions}

In this paper we studied self-accelerating solutions in the most general scalar-tensor theory compatible with the constraint that the speed of gravity and light are the same. This theory belongs to the class of degenerate higher order scalar-tensor theories that are free from Ostrogradsky instabilities despite the presence of higher order derivatives in the field equations. 

We found that scaling solutions, analogous to those found in covariant galileons, exist both in the matted dominated era and late time de Sitter phase, and the universe dynamically evolves from the former to the latter. Using these solutions, we computed the evolution of density perturbations. In a particular example studied in this paper, the density perturbation and lensing potential are enhanced compared to $\Lambda$CDM. There is an interesting possibility in this theory that the enhancement of the density perturbation and lensing potential could be compensated by the extra dumping term, depending on the choice of parameters. Finally, we showed that it is possible to evade the stringent constraints coming from the stellar structure and the orbital decay of Hulse-Taylor pulsar. 

A full scanning of the parameter space of the model is necessary in order to completely address the question whether this theory is a viable alternative to the $\Lambda$CDM model or not. This requires a modification of Einstein-Boltzmann codes to compute cosmological observables and perform parameter estimations by imposing, at the same time, the stability conditions derived in \cite{Langlois:2017mxy}. The theory considered in this paper is not included in the recent extensions of the Einstein-Boltmann codes \cite{Bellini:2017avd} and this deserves further investigations.  

\subsection*{Acknowledgments}
MC is supported by the Labex P2IO and the Enhanced Eurotalents Fellowship. KK is supported by the STFC grant ST/N000668/1.  The work of KK and MC has received funding from the European Research Council (ERC) under the European Union's Horizon 2020 research and innovation programme (grant agreement 646702 ``CosTesGrav"). 

\begin{appendix}

\section{Explicit form of the coefficients}

In this Appendix we provide the explicit expression for the coefficients used in the background, linear and non-linear equations.

\subsection{Background equations}
\label{background}

\bea
&& {\cal A} \equiv G \left(3-9 \B_1\right) - 6 \dot\j^2 G_X \,, \\
&& {\cal B} \equiv -\frac{3}{\dot\j} \left[2 \ddot\j \B_1 \left(2 \dot\j^2 G_X+G \left(3 \B_1-1\right)\right)+\dot\j^4
   G_{3X}\right] \,, \\
&& {\cal C} \equiv \dot\j^2 G_{2X}+\frac{G_2}{2}
   \left(3 \B_1+1\right) + \frac{3 \ddot\j \B_1}{\dot\j^2}
   \left[\ddot\j \B_1
   \left(G \left(1-3 \B_1\right)-2 \dot\j^2 G_X\right)-\dot\j^4 G_{3X}\right] \,, \\
&& {\cal D} \equiv \frac{18 \ddot\j
   G}{\dot\j^2} \left(2 \dot\j^2 \B_{1X}-3 B_1^2+\B_1\right) -\frac{48 \dot\j^2 \ddot\j G_X^2}{G} - 3
   \left[\dot\j^2 \left(3 G_{3X}-8
   \ddot\j G_{XX}\right)+4 \ddot\j G_X \left(6 \B_1+1\right)\right]  \\
&& {\cal E} \equiv
   \frac{6\dot\j G_X}{G}
   \left(G_2-4 \ddot\j \left(4
   \ddot\j G_X \B_1+\dot\j^2
   G_{3X}\right)\right) + \frac{36 \ddot\j^2 G \B_1}{\dot\j^3} \left(2
   \dot\j^2 \B_{1X}-3
   \B_1^2+\B_1\right)  \\
   && + \frac{6}{\dot\j}
   \left[-4 \ddot\j^2 \B_1 \left(G_X
   \left(6 \B_1+1\right)-2 \dot\j^2 G_{XX}\right)+\dot\j^2
   G_{2X}+3 G_2 \B_1+2 \dot\j^2 \ddot\j \left(\dot\j^2
   G_{3XX}-G_{3X} \left(3 \B_1+1\right)\right)\right] \,, \nb \\
&& {\cal F} \equiv \frac{3 G_2}{2 \dot\j^2G} \left[2
   \ddot\j \left(2 \dot\j^2 G_X \B_1+G \left(\B_1 \left(3 \B_1-1\right)-2 \dot\j^2 \B_{1X}\right)\right)+\dot\j^4 G_{3X}\right]+ \nb \\
    && \qquad \frac{\ddot\j}{\dot\j^4G} \left[-3 \dot\j^4
   \left(4 \ddot\j G_X \B_1+\dot\j^2
   G_{3X}\right)^2 +18 \ddot\j^2 G^2 \B_1^2 \left(2 \dot\j^2 \B_{1X}-3 \B_1^2+\B_1\right) \right] \nb \\
   && \qquad + \frac{\ddot\j}{\dot\j^2}
   \left[-12 \ddot\j^2 \B_1^2 \left(G_X
   \left(6 \B_1+1\right)-2 \dot\j^2 G_{XX}\right) -4 \dot\j^4
   G_{2XX}+2 \dot\j^2
   G_{2X} \left(1-3 \B_1\right) \right. \nb \\
   && \qquad \left. -3 \dot\j^2 \ddot\j \B_1
   \left(G_{3X} \left(9 \B_1+4\right)-4 \dot\j^2
   G_{3XX}\right)\right] \,.
\eea

\subsection{Non-linear equations}
\label{perturbations}

\bea
&& {\cal G}_T \equiv 2 \dot\j^2 \left(2  G_X + \frac{G}{\dot\j^2} \right) \,, \qquad
{\cal G}_{T \Phi} \equiv 2 \B_1\dot\j^2 \left[ 4 G_X + \frac{G}{\dot\j^2} \left(\B_1+2\right) \right] \,,  \\
&& a_2 \equiv 8 \left(G_X+ \frac{G \B_1}{\dot\j^2}\right) \left(H \dot\j+\ddot\j \B_1\right) + \dot\j^2 G_{3X} \,, \qquad
a_{2}^t \equiv -2 \B_1 \dot\j \left[4 G_X+\frac{G}{\dot\j^2} \left(\B_1+1\right)\right] \,, \\
&& b_2^{a} \equiv - \frac{2}{a^2} \left(G_X+\frac{G \B_1}{\dot\j^2}\right) \,, \qquad
b_2^{b} \equiv \frac{2}{a^2} \left[G_X \left(4 \B_1+1\right)+\frac{G \B_1}{\dot\j^2} \left(\B_1+2\right)\right] \,, \\
&& b_2^{c} \equiv \frac{2 \B_1}{a^2} \left[4 G_X+\frac{G}{\dot\j^2} \left(\B_1+1\right)\right] \,, \\
&& {\cal F}_T \equiv 2 G \,, \qquad
a_1^t \equiv 4 \dot\j G_X \,, \qquad
b_1 \equiv - \frac{4 G_X}{a^2} \,, \\
&& a_1 \equiv 8 \left[G_X \left(H \dot\j+\ddot\j\right)-2 \dot\j^2 \ddot\j G_{XX}\right] \,, \qquad
b_3 \equiv - \frac{4}{a^2} \left[2 G_X \left(2 \B_1+1\right)+\frac{G \B_1}{\dot\j^2} \left(\B_1+3\right)\right] \,, \\
&& a_0^{tt} \equiv -4 \B_1 \left(4 G_X+\frac{G \B_1}{\dot\j^2}\right) \,,\qquad
b_0^t \equiv \frac{4 \B_1}{a^2 \dot\j} \left(4 G_X+\frac{G \B_1}{\dot\j^2}\right) \,, \\ 
&& c_0^a \equiv \frac{4}{a^4 \dot\j^2} \left(G_X+\frac{G \B_1}{\dot\j^2}\right) \,,\qquad
c_0^c \equiv -\frac{4 \B_1}{a^4 \dot\j^2} \left(4 G_X+\frac{G \B_1}{\dot\j^2}\right) \,, \\
&& c_0^b \equiv -\frac{4}{a^4 \dot\j^2} \left[G_X \left(4\B_1+3\right)+\frac{G \B_1}{\dot\j^2} \left(\B_1+3\right)\right] \,, \\
&& a_0 \equiv \frac{4}{\dot\j^4} \left[-8 \dot\j^4 \ddot\j G_{XX} \left(H \dot\j+\ddot\j
   \B_1\right)+\dot\j^2
   G_X \left(4 \dot\j
   \left(\dot\j \left(\dot H-2 \ddot\j^2 \B_{1X}\right)+\dddot\j \B_1\right)-4 H \dot\j
   \ddot\j \left(\B_1-1\right) \right.\right. \nb \\
   &&\left.\left. \qquad +7 H^2 \dot\j^2-11 \ddot\j^2 \B_1^2\right)+G \left(\B_1 \left(7 \dot H \dot\j^2+H \dot\j \ddot\j \left(13
   \B_1-4\right)+13 H^2
   \dot\j^2+\B_1 \left(7
   \dddot\j \dot\j-11 \ddot\j^2\right)\right) \right.\right. \nb \\
   && \left.\left. \qquad -2 \dot\j^2 \ddot\j \B_{1X} \left(4 H \dot\j+11
   \ddot\j \B_1\right)\right)\right] 
   + 2 \left[-G_{2X}+4 H \dot\j G_{3X}+2 \ddot\j
   \left(G_{3X}-\dot\j^2
   G_{3XX}\right)\right] \,, \\
&& a_0^t \equiv \frac{8}{\dot\j} \left[\ddot\j \left(4 \dot\j^2 \B_1 G_{XX}+G_X \left(4 \dot\j^2 \B_{1X}+\B_1^2\right)\right)-2 H \dot\j G_X \B_1\right] \nb \\
&& \qquad +\frac{4 G \B_1}{\dot\j^3} \left[2 \ddot\j \left(2
   \dot\j^2 \B_{1X}+\B_1\right)-H \dot\j \B_1\right] \,, \\
&& a_3 \equiv -\frac{4}{\dot\j^2} \left[8 \dot\j^4 \ddot\j \B_1 G_{XX}+\dot\j^2 G_X
   \left(2 \ddot\j \left(4 \dot\j^2 \B_{1X}+\left(\B_1-3\right) \B_1\right)-4
   H \dot\j \left(\B_1+1\right)\right) \nb \right. \\
   && \left. \qquad +G \left(\B_1 \left(\ddot\j \left(1-3
   \B_1\right)-H \dot\j \left(\B_1+5\right)\right)+2
   \dot\j^2 \ddot\j \left(2 \B_1+1\right) \B_{1X}\right)\right] + 2 \dot\j^2 G_{3X} \,,  \\
&& b_0^a \equiv -\frac{4}{a^2 \dot\j^4} \left[-2 \dot\j^4 \ddot\j G_{XX}+\dot\j^2 G_X
   \left(3 H \dot\j+\ddot\j \left(2 \B_1+1\right)\right) \nb \right. \\
   && \left. \qquad +G \left(\B_1 \left(3
   H \dot\j+\ddot\j \left(4 \B_1-1\right)\right)-2 \dot\j^2
   \ddot\j \B_{1X}\right)\right] \,, \\
&& b_0^b \equiv \frac{4}{a^2 \dot\j^4} \left[-2 \dot\j^4 \ddot\j \left(4 \B_1+1\right) G_{XX}+\dot\j^2 G_X
   \left(H \dot\j \left(3-4 \B_1\right)-\ddot\j \left(8 \dot\j^2
   \B_{1X}+2 \B_1 \left(\B_1+1\right)-1\right)\right) \right. \nb \\
   && \left. \qquad +G \left(-\B_1 \left(H
   \dot\j \left(\B_1-3\right)-\ddot\j \left(\B_1-1\right)\right)-2 \dot\j^2
   \ddot\j \left(2 \B_1+1\right) \B_{1X}\right)\right] \,, \\
&& b_0^c \equiv -\frac{4}{a^2\dot\j^4} \left[8 \dot\j^4 \ddot\j \B_1 G_{XX}+2
   \dot\j^2 G_X \left(\B_1 \left(2 H \dot\j+\ddot\j \left(\B_1+2\right)\right)+4 \dot\j^2 \ddot\j
   \B_{1X}\right) \right. \nb \\
   && \left.\qquad  +G \B_1 \left(\B_1 \left(H \dot\j+3
   \ddot\j\right)+4 \dot\j^2 \ddot\j
   \B_{1X}\right)\right]  \,. 
\eea

\subsection{Linear equations}
\label{linear}

\bea
&& \gamma_1 \equiv a_0-\frac{a_2 \left( a_1
   \G_T + a_3 \F_T\right)+2 \dot a_2 (a_1^t\G_T - a_2^t \F_T)}
   {\F_T \G_{T\Phi}+\G_T^2}  \nb \\
&&\qquad +\frac{2
   a_2}{\left(\F_T \G_{T\Phi}+\G_T^2\right)^2} \left[a_1^t
   \left(\G_T \left(\G_{T\Phi} \dot\F_T+\F_T
   \dot\G_{T\Phi}\right)-\F_T
   \G_{T\Phi}
   \dot\G_T+\G_T^2
   \dot\G_T\right) \right. \nb \\
&& \left. \qquad\qquad\qquad \qquad\qquad+a_2^t
   \left(\G_T^2 \dot\F_T-2
   \F_T \G_T
   \dot\G_T-\F_T^2
   \dot\G_{T\Phi}\right)\right] \,, \\[2ex]
&& \gamma_2 \equiv \frac{ a_1\G_T +a_3
   \F_T }{\F_T \G_{T\Phi}+\G_T^2}
   +\frac{2}{\left(\F_T \G_{T\Phi}+\G_T^2\right)^2}
   \left[a_2^t \left(-\G_T^2
   \dot\F_T+\F_T
   \G_T \left(2
   \dot\G_T+H\G_T
   \right)+\F_T^2
   \left(\dot\G_{T\Phi}+H\G_{T\Phi}
   \right)\right) \right. \nb \\
   && \qquad \left. -a_1^t
   \left(\G_T \left(\G_{T\Phi}
   \left(\dot\F_T+H\F_T
   \right)+\F_T
   \dot\G_{T\Phi}\right)-\F_T
   \G_{T\Phi}
   \dot\G_T+\G_T^2
   \dot\G_T+H\G_T^3
   \right)\right] \,, \\[2ex]
&& \gamma_3 \equiv \frac{2 \left( a_2^t \F_T-
   a_1^t
   \G_T\right)}{\F_T \G_{T\Phi}+\G_T^2} \,.
\eea

\bea
&& D \equiv \g_1\left(2\g_1 + \rho \g_3^2 \right)
   \left(\F_T \G_{T\Phi}+\G_T^2\right) \,, \\
&& \mu_\Phi \equiv \frac{1}{D} \left[ 2\g_1\left( \g_1+a_2\g_2 \right)\F_T - \g_3 
   \left(\g_1
   \left(H \g_2-\dot\g_2\right)+\g_2
   \dot\g_1\right)\left(\F_T \G_{T\Phi}+\G_T^2\right)\right] \,,\\
&& \nu_\Phi \equiv \frac{\g_3}{D} \left[2a_2
   \g_1
   \F_T -  
   \left(\g_1 \left(3
   H \g_3+\g_2-\dot\g_3\right)+\g_3
   \dot\g_1\right)\left(\F_T \G_{T\Phi}+\G_T^2\right) \right] \,, \nb \\
&& \mu_\Psi \equiv \frac{2}{D}\left[ 
   (\g_1+a_2
   \g_2)\left(\g_1 \G_T - \rho \g_3 a_1^t \right)
   -\left(\g_1
   \left(H \g_2-\dot\g_2\right)+\g_2
   \dot\g_1\right)(a_1^t \G_{T\Phi}+a_2^t \G_T) \right] \,, \\
&& \nu_\Psi \equiv \frac{2}{D} \left[a_2 \g_3
   (\g_1\G_T-\rho \g_3 a_1^t )-
   \left(\g_1 \left(3
   H \g_3+\g_2-\dot\g_3\right)+\g_3
   \dot\g_1\right)(a_1^t \G_{T\Phi}+a_2^t \G_T)\right] \,.
\eea

\end{appendix}

\end{document}